\documentclass[structabstract]{aa}

\usepackage{txfonts}
\usepackage{graphicx}
\usepackage{pstricks}

\usepackage{rotating}

\usepackage{hyperref}
\hypersetup{
    bookmarks=true,         
    unicode=false,          
    pdftoolbar=true,        
    pdfmenubar=true,        
    pdffitwindow=false,     
    pdftitle={EL61 Family},
    pdfauthor={}, 
    pdfsubject={Planetary Science}, 
    pdfkeywords={},         
    pdfnewwindow=true,      
    colorlinks=true,        
    linkcolor=gray,         
    citecolor=blue,         
    filecolor=gray,         
    urlcolor=gray           
}

\usepackage[authoryear,square]{natbib}

\begin{document}

\title{Characterisation of candidate members\\ of (136108) Haumea's family
  \thanks{Based on observations
    collected at the European Southern Observatory, La Silla \& Paranal, Chile -
    \href{http://archive.eso.org/wdb/wdb/eso/sched_rep_arc/query?progid=81.C-0544}%
         {81.C-0544}
  \& 
    \href{http://archive.eso.org/wdb/wdb/eso/sched_rep_arc/query?progid=82.C-0306}%
         {82.C-0306}
}}


\author{Colin Snodgrass\inst{1,2}
  \and Beno\^{i}t Carry\inst{1,3}
  \and Christophe Dumas\inst{1}
  \and Olivier Hainaut\inst{4}
}

\offprints{C. Snodgrass, \email{snodgrass@mps.mpg.de}}

\institute{European Southern Observatory, Alonso de C\'{o}rdova 3107,
  Vitacura, Casilla 19001, Santiago de Chile, Chile
  \and Max Planck Institute for Solar System Research, Max-Planck-Strasse 2,
  37191 Katlenburg-Lindau, Germany
  \and LESIA, Observatoire de Paris-Meudon,
  5 place Jules Janssen, 92195 Meudon Cedex, France
  \and European Southern Observatory,
  Karl-Schwarzschild-Strasse 2,
  D-85748 Garching bei M\"{u}nchen, Germany
}

\date{Received  / Accepted }

\abstract
{Ragozzine \& Brown [2007] presented a list of candidate members of the first collisional family to be found among the trans-Neptunian Objects (TNOs), the one associated with (136108) Haumea (2003 EL$_{61}$).}
{We aim to identify which of the candidate members of the Haumea collisional family are true members, by searching for water ice on their surfaces. We also attempt to test the theory that the family members are made of almost pure water ice by using optical light-curves to constrain their densities.}
{We use optical and near-infrared photometry to identify water ice, in particular using the $(J - H_S)$ colour as a sensitive measure of the absorption feature at 1.6 $\mu m$. We use the $CH_4$ filter of the new Hawk-I instrument at the VLT as a short $H$-band ($H_S$) for this as it is more sensitive to the water ice feature than the usual $H$ filter.}
{We report colours for 22 candidate family members, including NIR colours for 15. We confirm that 2003 SQ$_{317}$ and 2005 CB$_{79}$ are family members, bringing the total number of confirmed family members to 10. We reject 8 candidates as having no water ice absorption based on our Hawk-I measurements, and 5 more based on their optical colours. The combination of the large proportion of rejected candidates and time lost to weather prevent us from putting strong constraints on the density of the family members based on the light-curves obtained so far; we can still say that none of the family members (except Haumea) require a large density to explain their light-curve.}
{}


\maketitle

\section{Introduction}

The trans-Neptunian object (TNO) (136108) Haumea (2003 EL$_{61}$) was discovered by \citet{2005-IAUC-8577-Santos-Sanz} and quickly attracted a lot of attention as a highly unusual body. It is one of the largest TNOs \citep{2006-ApJ-639-Rabinowitz,2008ssbn.book..161S} and yet is a fast rotator (period $\sim 3.9$ hours) with a highly elongated shape \citep{2006-ApJ-639-Rabinowitz}. Its surface was shown to be dominated by water ice by Near Infra-Red (NIR) spectroscopy \citep{2007-AJ-133-Tegler,2007-ApJ-655-Trujillo,2007-AA-466-Merlin,2009-AA-496-Pinilla-Alonso}, yet has a high density of 2.5-3.3 g cm$^{-3}$ \citep{2006-ApJ-639-Rabinowitz}. It was found to have two satellites \citep{2005-ApJ-632-Brown,2006-ApJ-639-Brown}, which also have water ice surfaces \citep{2006-ApJ-640-Barkume,2009-ApJ-695-Fraser}. \citet{2008-AJ-135-Lacerda} found that Haumea presents hemispherical colour heterogeneity, with a dark red `spot' on one side, using high precision photometry.

\citet{2006-ApJ-639-Brown} and \citet{2006-ApJ-640-Barkume} postulated that the density, shape and water ice surface could be explained by a large collision early in the history of the Solar System. \citet{2007-Nature-446-Brown} then identified a family of 6 TNOs (1995 SM$_{55}$, 1996 TO$_{66}$, 2002 TX$_{300}$, 2003 OP$_{32}$ and 2005 RR$_{43}$), in addition to Haumea and its satellites, with orbits that could be linked to Haumea and water ice surfaces, which were also attributed to coming from this massive collision. This theory required that the proto-Haumea was a very large body (radius $\sim 830$ km) that had already differentiated early in the formation of the Solar System, and that the collision stripped nearly all of the outer (water ice) mantle ($\sim 20\%$ of the total mass of the original body). This left the dense core as Haumea with a thin coating of water ice and created a family of re-accumulated lumps of almost pure water ice. \citet{2007-AJ-134-Ragozzine} find that the collision must have taken place in the early Solar System (with an age of at least 1 Gyr), although the lack of weathering on the surfaces may imply young bodies \citep{2008-AJ-136-Rabinowitz}. The existence of such a family has implications for the dynamics of the Kuiper Belt \citep{2008-AJ-136-Levison}.

\citet{2007-AJ-134-Ragozzine} performed a dynamical study and identified two further family members (2003 UZ$_{117}$ and 1999 OY$_3$) with strong dynamical links to the family and colours consistent with water ice, and also published a list of candidate family members that had orbital elements consistent with this dynamical family, totalling 35 objects including the known members. Most of these candidates lacked the NIR spectra that could identify water ice on their surfaces though, so they remained only potential family members. The diffusion time and interaction with resonances make it possible for interlopers to appear close to the family dynamically, so it is essential to have both dynamical and physical properties characterisation to confirm family membership \citep{2002-AsteroidsIII-5.1-Cellino}. Some could be ruled out by either existing NIR spectra (Makemake has a methane ice surface; \citet{2007-AA-471-Dumas,2007-AJ-133-Brown}) or by very red optical colours (1996 RQ$_{20}$, 1999 CD$_{158}$, 1999 KR$_{16}$, 2002 AW$_{197}$, 2002 GH$_{32}$; see table \ref{tab: colors} for references) or a strong red slope in optical spectra (2005 UQ$_{513}$; \citet{2008-AA-489-Pinilla-Alonso}). \citet{2008-ApJ-684-Schaller} subsequently published NIR spectra which confirmed 2003 UZ$_{117}$ and 2005 CB$_{79}$ as family members, and rejected 2004 SB$_{60}$. We observed 13 of the 18 remaining candidate objects (along with some of the already characterised objects) with the goal of providing this physical information, to identify those with water ice surfaces and also to test the idea that these family members could be made of nearly pure water ice. We describe our observations, the results from them, and their implications in the following sections.


\section{Observations and data reduction}

\begin{table}[h!]
  \caption{%
    Observational circumstances.
     }
  \label{tab: obs-log}
\begin{center}
\begin{tabular}{rl cccc p{0.15em}p{0.15em}p{0.3em}p{0.15em}}
  \hline\hline
  \multicolumn{2}{c}{Object} & $r$$^a$ & $\Delta$$^b$ & $\alpha$$^c$ & Run$^d$ & \multicolumn{4}{c}{Epochs$^e$} \\
  (\#) & (Designation) & (AU) & (AU) & (\degr) & & B & V & R & i\\
  \hline
& 1996 RQ 20   & 39.6 & 39.0 & 1.1  & C & & & 4 & \\ 
 20161 & 1996 TR 66   & 40.3 & 40.0 & 1.4 & E & 2 & 2 & 2 & 2 \\ 
& 1998 HL 151   & 38.9 & 38.2 & 1.0 & A & 2 & 2 & 2 & 2 \\ 
181855 & 1998 WT 31   & 38.0 & 37.3 & 1.0 & E & 2 & 2 & 10 & 2 \\ 
& 1999 CD 158  & 47.6 & 46.5 & 0.6 & E & 1 & 1 & 24 & 1 \\ 
40314 & 1999 KR 16    & 36.3 & 35.6 & 1.2& B \\
& 1999 OH 4     & 39.1 & 39.6 & 1.3 & A & 1 & 1 & 1 & 1 \\ 
& ~~~~~"     & 39.1 & 38.2 & 0.6  & C & 2 & 2 & 2 & 2 \\ 
& 1999 OK 4     & 46.4 & 45.8 & 1.1 & A & 1 & 1 & 1 & 1 \\ 
 86047 & 1999 OY 3     & 40.1 & 39.7 & 1.3 & A & & & 11 & \\ 
 "~~ & ~~~~~"     & 40.2 & 39.5 & 1.1& B \\
 "~~ & ~~~~~"   & 40.2 & 39.4 & 0.8 & D \\
 86177 & 1999 RY 215  & 35.8 & 34.8 & 0.2 & C & & & 21 & \\ 
 "~~ & ~~~~~"  & 35.8 & 34.8 & 0.3 & D \\
& 2000 CG 105  & 46.8 & 46.1 & 0.8 & E & 2 & 2 & 22 & 2 \\ 
130391 & 2000 JG 81    & 34.8 & 33.8 & 0.5  & A & 1 & 1 & 1 & 1 \\ 
& 2001 FU 172   & 31.8 & 30.9 & 1.0 & A & 1 & 1 & 1 & 1 \\ 
& 2001 QC 298  & 40.6 & 39.6 & 0.3 & C & & & 17 & \\ 
&  ~~~~~"   & 40.6 & 39.6 & 0.2 & D \\
55565 & 2002 AW 197  & 46.6 & 45.8 & 0.7 & F \\
& 2002 GH 32    & 43.1 & 42.2 & 0.7 & A &  3 & 3 & 18 & 3 \\ 
 & ~~~~~"       & 43.1 & 42.4 & 1.0& B \\
 55636 & 2002 TX 300   & 41.4 & 40.6 & 0.8 & D \\
136108 & Haumea        & 51.1 & 50.6 & 1.0 & A & 1 & 1 & 1 & 1 \\ 
"~~ & ~~~~~"         & 51.1 & 50.8 & 1.1& B \\
"~~ & ~~~~~"       & 51.1 & 51.1 & 1.1 & F \\
& 2003 HA 57    & 32.7 & 32.0 & 1.3  & A & 1 & 1 & 1 & 1 \\ 
& ~~~~~"      & 32.7 & 32.2 & 1.6& B \\
& 2003 HX 56    & 46.5 & 45.9 & 1.0  & A & 2 & 2 & 2 & 2 \\ 
120178 & 2003 OP 32    & 41.4 & 40.6 & 0.6 & D \\
& 2003 QX 91   & 33.6 & 32.6 & 0.5& C & & & 4 & \\ 
& 2003 SQ 317  & 39.3 & 38.3 & 0.6 & C & & & 15 & \\ 
&  ~~~~~"    & 39.3 & 38.3 & 0.4 & D \\
& 2003 TH 58   & 36.0 & 35.1 & 0.5 & E & 2 & 2 & 23 & 2 \\ 
&  ~~~~~"    & 36.0 & 35.1 & 0.7 & F \\
136199 & Eris          & 96.7 & 95.9 & 0.4 & D \\
& 2003 UZ 117   & 39.4 & 38.9 & 1.3 & D \\
& 2004 PT 107   & 38.3 & 37.9 & 1.4  & A &  4 & 4 & 24 & 4 \\ 
& ~~~~~"    & 38.3 & 37.7 & 1.3& B \\
&  ~~~~~"   & 38.3 & 37.4 & 0.7 & D \\
120347 & 2004 SB 60   & 44.0 & 43.1 & 0.6& C & & & 16 & \\ 
& 2005 CB 79   & 40.1 & 39.3 & 0.9 & E & 1 & 1 & 2 & 1 \\ 
&  ~~~~~"    & 40.0 & 39.2 & 0.8 & F \\
& 2005 GE 187   & 30.8 & 29.9 & 0.9 & A & 3 & 3 & 33 & 3 \\ 
&  ~~~~~"     & 30.8 & 30.1 & 1.3 & B \\
&  ~~~~~"   & 30.8 & 31.1 & 1.7 & C & & & 17 & \\ 
&  ~~~~~"    & 30.8 & 31.3 & 1.6 & D \\
202421 & 2005 UQ 513  & 48.8 & 48.1 & 0.8 & C & & & 10 & \\ 
"~~ & ~~~~~"   & 48.8 & 48.0 & 0.7 & D \\
  \hline
\end{tabular}
\end{center}
Notes:\\
$^a$ Heliocentric distance; $^b$ Geocentric distance; $^c$ Phase angle.\\
$^d$ Runs: A = 2008 June 3rd -- 5th, EFOSC2;
  B = 2008 June 17th, Hawk-I;
  C = 2008 August 30th -- September 1st, EFOSC2;
  D = 2008 September 9th, Hawk-I;
  E = 2008 December 29th -- 31st, EFOSC2;
  F = 2009 January 4th, Hawk-I.\\
$^e$ Number of epochs observed in each filter (for EFOSC2 runs).\\

\end{table}

\begin{table}
  \caption{%
    Filters used in this study.}
  \label{tab: obs-filter}
\begin{center}
\begin{tabular}{cccc}
  \hline\hline
  Filter & Instrument & $\lambda_c$ & $\Delta \lambda$ \\
  & & $\mu$m & $\mu$m \\
  \hline
  B   & EFOSC2 & 0.440 & 0.094 \\
  V   & EFOSC2 & 0.548 & 0.113 \\
  R   & EFOSC2 & 0.643 & 0.165 \\
  i   & EFOSC2 & 0.793 & 0.126 \\
  J   & Hawk-I & 1.258 & 0.154 \\
  H$_S$ (CH4) & Hawk-I & 1.575 & 0.112 \\
  \hline
\end{tabular}
\end{center}
Notes:\\
$\lambda_c$ = Central Wavelength, $\Delta
    \lambda$ = Bandwidth.
\end{table}

The best method to test for water ice on the surface of a Solar System body is through NIR spectroscopy, as water ice has strong absorption bands at $\sim 1.6$ and $\sim 2.0$ $\mu$m, but this is only possible for the brightest TNOs ($K \lesssim 18$). Still, it is possible to get an indication of the presence or absence of water ice for fainter bodies using photometry, which can be performed on smaller (fainter) TNOs.

We conducted the observations at the European Southern
  Observatory (program IDs:
  \href{http://archive.eso.org/wdb/wdb/eso/sched_rep_arc/query?progid=81.C-0544}%
       {81.C-0544}
  \& 
  \href{http://archive.eso.org/wdb/wdb/eso/sched_rep_arc/query?progid=82.C-0306}%
       {82.C-0306}), on both the La Silla and Paranal (VLT) sites.
  Observations in the visible wavelengths ($BVRi$ filters) were performed 
  using the EFOSC2 instrument
  \citep{1984-Msngr-38-Buzzoni} mounted on the NTT 
  \citep[since April 2008;][]{2008-Msngr-132-Snodgrass}. This is a focal reducing imager and spectrograph with a single CCD. 
  The near-infrared observations ($J$, $CH_4$ bands) were performed using
  the newly commissioned wide-field camera Hawk-I
  \citep{2004-SPIE-5492-Pirard, 2006-SPIE-6269-Casali}. We had three observing runs scheduled with each instrument, as detailed in table \ref{tab: obs-log}. This table lists all objects we attempted to observe, although not all were detected and some time was lost to poor weather conditions. In particular the June 17th Hawk-I run (run B) was very badly affected by clouds, with only 1999 KR$_{16}$ reliably detected in both bands. Exposure times were generally 300-600 seconds in the optical, while in the NIR we took sequences of $J$-$CH_4$-$J$ to give an average $J$ magnitude at the time of the $CH_4$ observations, and to confirm identification of the object based on its motion between the two sets of $J$-band images. The $CH_4$ filter observations took the largest part of the time; between 15 minutes for the brightest objects to a few hours for the faintest ones, each split into short individual exposures and dithered due to the bright NIR sky. Note that due to the long effective exposure times any variation (due to shape or albedo variation across the surface) is smeared out, and cannot be detected in our NIR data.

The advantage of using Hawk-I is that the $CH_4$ band filter is a medium width filter with a wavelength range that is entirely within the broad water ice absorption between 1.4 and 1.75 $\mu$m. The standard $H$-band is broader and covers a range that is part in and part out of this band\footnote{See 
\href{http://www.eso.org/sci/facilities/paranal/instruments/hawki/inst}%
{http://www.eso.org/sci/facilities/paranal/instruments/hawki/inst}
for transmission curves.}
. We therefore use the $CH_4$ filter as a short $H$ filter (henceforth $H_S$) which gives a colour measurement $(J-H_S)$ that is very sensitive to water ice absorption. All of the filters used in this work are listed in table \ref{tab: obs-filter}.

The data were reduced in the normal manner (bias subtraction, flat fielding, sky subtraction etc.~as appropriate). For the EFOSC2 data the objects were generally visible in individual frames and aperture photometry was performed directly on each, using the optimum aperture based on the measured stellar FWHM in each frame and an average aperture correction measured using the field stars \citep[see][]{2005-AA-444-Snodgrass}. Where multiple epochs were obtained we then report a weighted mean magnitude. This approach allowed us to look for variation in the $R$-band magnitude for those objects where we obtained a light-curve. For fainter objects the images were shifted based on the predicted motion of each object and combined to give a deep image per filter. We also produced equivalent combined images of the star fields (no shifts) in which we could measure the brightness of field stars for photometric calibration. For Hawk-I all data were shifted and combined as the individual exposures were short because of the high sky background in the NIR.

The EFOSC2 data were calibrated in the normal way, via observations of standard stars from the \citet{1992-AJ-104-Landolt} catalogue. The EFOSC2 $i$-band data was calibrated directly onto the Landolt scale; this filter is very close to the standard Cousins $I$-band used by Landolt. Data from non-photometric nights were calibrated via observation of the same fields on later photometric nights, to calibrate the field stars as secondary standard stars.

\begin{figure}
   \centering
   \includegraphics[angle=-90,width=0.45\textwidth]{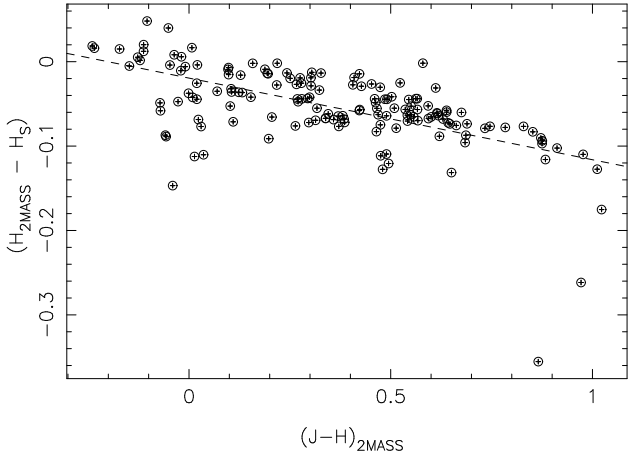} 
   \caption{Theoretical difference between 2MASS $H$ and Hawk-I $H_S$ for different stellar spectra, as a function of 2MASS $(J-H)$.}
   \label{fig:Hs_cal}
\end{figure}

Calibration of the Hawk-I data was a more involved process as it contained the non-standard filter $H_S$. The $J$ and $H$ band magnitudes of all available stars in each field were taken from the 2MASS point source catalogue \citep{2006-AJ-131-Skrutskie}. We then generated theoretical colours $(H_{2M} - H_S)$ for stars of all spectral types (O-M) by convolving the response of the 2MASS $H$ and the Hawk-I $H_S$ with spectra from the libraries of \citet{1998-PASP-110-Pickles} and \citet{2004-ApJS-151-Ivanov}\footnote{These libraries can be downloaded from the ESO web pages at \href{http://www.eso.org/sci/observing/tools/standards/IR_spectral_library_new/}{http://www.eso.org/sci/observing/tools/standards/IR\_spectral\_library\_new/}}. For stars the resulting difference is linearly related to the 2MASS $(J-H)$ colour (Fig.~\ref{fig:Hs_cal}):
\begin{equation}
(H_{2M} - H_S) = -0.097(J - H)_{2M} - 0.019.
\end{equation}
We used this relation to generate the expected colour, and therefore $H_S$ magnitude, for each 2MASS star in each field, which were then used to give the calibrated $H_S$ magnitude for the TNOs. We also used the same approach to derive the colour term for the difference between 2MASS and Hawk-I $J$ bands, and found that the Hawk-I $J$ does not significantly differ from the 2MASS band, as expected. We note that the spectral types further from the linear trend fall into two groups; those below the trend at $(J-H)_\textrm{2M} \approx 0$ are B stars that do not feature in our NIR images, while the `tail' that curves away from the line at the red end is made up of M giants, with M8-10 being significant away from the linear relation. These are separable from the rest of the sample though as giants have a very red 2MASS $(J-K)$ colour; \citet{2003-ApJ-593-Brown} show that stars with $(J-K) \ge 0.5$ are most likely giants, while we find that using limit of $(J-K) \le 1.26$ removes the M8-10iii stars that do not fit the linear trend while keeping other stars. Having said this, we note that the exclusion or inclusion of these stars made no significant difference to our calibration as there were very few late M giant stars within our sample. The colours of the 2MASS stars in the fields observed were approximately normally distributed around a mean $(J-H)_\textrm{2M} = 0.6$ with a standard deviation of 0.2.


\section{Colours}
\label{colours_section}

\begin{table*}
\caption{Photometry. Mean apparent magnitudes for each object at each epoch.}
  \label{tab: phot}
\begin{center}
\begin{tabular}{rl c cccccc c}
  \hline\hline
  \multicolumn{2}{c}{Object} & Run$^a$ & $B$  & $V$ & $R$ & $I$ & $J$ & $H_S$ & $\Delta m_R$$^b$ \\
  \hline
 & 1996 RQ 20  & C   &         --       &         --       & 22.95 $\pm$ 0.05 &         --       &         --       &        --         & --\\
 & 1998 HL 151 & A   & 25.37 $\pm$ 0.28 & 24.25 $\pm$ 0.12 & 23.87 $\pm$ 0.13 & 23.08 $\pm$ 0.17 &         --       &        --         & --\\
181855 & 1998 WT 31  & E   & 24.52 $\pm$ 0.15 & 23.81 $\pm$ 0.11 & 23.24 $\pm$ 0.06 & 22.69 $\pm$ 0.14 &         --       &        --       & \textless 0.1\\
 & 1999 CD 158 & E   & 23.08 $\pm$ 0.07 & 22.31 $\pm$ 0.06 & 21.68 $\pm$ 0.01 & 21.20 $\pm$ 0.06 &         --       &        --       & 0.6\\
40314 & 1999 KR 16  & B   &         --       &         --       &         --       &         --       & 20.02 $\pm$ 0.07 & 19.47 $\pm$ 0.10   & --\\ 
 & 1999 OH 4   & C   & 25.01 $\pm$ 0.22 & 22.39 $\pm$ 0.09 & 22.19 $\pm$ 0.10 & 21.76 $\pm$ 0.32 &         --       &        --         & --\\
86047 & 1999 OY 3   & A,D &         --       &         --       & 22.26 $\pm$ 0.03 &         --       & 21.78 $\pm$ 0.10 & 22.04 $\pm$ 0.35 & \textless 0.1\\
86177 & 1999 RY 215 & C,D &         --       &         --       & 22.16 $\pm$ 0.01 &         --       & 21.19 $\pm$ 0.14 & 20.66 $\pm$ 0.17 & \textless 0.1\\
 & 2000 CG 105 & E   & 24.14 $\pm$ 0.09 & 22.60 $\pm$ 0.04 & 22.62 $\pm$ 0.02 & 22.52 $\pm$ 0.09 &         --       &        --       & 0.45\\
 & 2001 FU 172 & A   & 26.71 $\pm$ 1.52 & 23.80 $\pm$ 0.15 & 23.13 $\pm$ 0.12 & 22.66 $\pm$ 0.19 &         --       &        --         & --\\
 & 2001 QC 298 & C,D &         --       &         --       & 22.18 $\pm$ 0.03 &         --       & 21.16 $\pm$ 0.08 & 20.65 $\pm$ 0.12 & 0.4\\
55565 & 2002 AW 197 & F   &         --       &         --       &         --       &         --       & 18.50 $\pm$ 0.05 & 18.11 $\pm$ 0.06  & --\\
 & 2002 GH 32  & A   & 23.91 $\pm$ 0.09 & 21.87 $\pm$ 0.05 & 21.87 $\pm$ 0.02 & 19.96 $\pm$ 0.09 &         --       &        --         & 0.75\\
55636 & 2002 TX 300 & D   &         --       &         --       &         --       &         --       & 18.67 $\pm$ 0.07 & 19.14 $\pm$ 0.10  & --\\
136108 & Haumea      & F   &         --       &         --       &         --       &         --       & 16.46 $\pm$ 0.07 & 17.06 $\pm$ 0.08  & --\\
 & 2003 HX 56  & A   & 25.25 $\pm$ 0.36 & 24.03 $\pm$ 0.16 & 23.68 $\pm$ 0.16 & 23.42 $\pm$ 0.43 &         --       &        --       & --\\
120178 & 2003 OP 32  & D   &         --       &         --       &         --       &         --       & 19.08 $\pm$ 0.05 & 19.58 $\pm$ 0.06 & --\\
 & 2003 QX 91  & C   &         --       &         --       & 23.66 $\pm$ 0.12 &         --       &         --       &        --       & --\\
 & 2003 SQ 317 & C,D &         --       &         --       & 22.05 $\pm$ 0.02 &         --       & 21.59 $\pm$ 0.05 & 22.04 $\pm$ 0.19 & 1.0 \\
 & 2003 TH 58  & E   & 23.50 $\pm$ 0.05 & 22.89 $\pm$ 0.04 & 22.51 $\pm$ 0.02 & 22.03 $\pm$ 0.04 & 21.73 $\pm$ 0.09 & 20.45 $\pm$ 0.18 & \textless 0.1\\
 & 2003 UZ 117 & D   &         --       &         --       &         --       &         --       & 20.24 $\pm$ 0.07 & 20.86 $\pm$ 0.10 & -- \\ 
 & 2004 PT 107 & A,D &         --       &         --       & 21.66 $\pm$ 0.01 &         --       & 20.41 $\pm$ 0.14 & 19.87 $\pm$ 0.18 & 0.05\\
120347 & 2004 SB 60  & C   &         --       &         --       & 20.21 $\pm$ 0.01 &         --       &         --       &        --       & 0.2\\
 & 2005 CB 79  & E,F & 21.45 $\pm$ 0.02 & 20.71 $\pm$ 0.03 & 20.36 $\pm$ 0.02 & 19.98 $\pm$ 0.03 & 19.67 $\pm$ 0.07 & 20.18 $\pm$ 0.16 & --\\
 & 2005 GE 187 & A   & 23.76 $\pm$ 0.10 & 22.78 $\pm$ 0.09 & 22.02 $\pm$ 0.01 & 21.47 $\pm$ 0.11 &         --       &        --       & \textless 0.1\\
 &  ~~~~~~~" & C,D &         --       &         --       & 22.13 $\pm$ 0.03 &         --       & 20.84 $\pm$ 0.08 & 20.18 $\pm$ 0.12 & 0.5\\
202421 & 2005 UQ 513 & C,D &         --       &         --       & 20.30 $\pm$ 0.01 &         --       & 18.89 $\pm$ 0.07 & 18.59 $\pm$ 0.10 & 0.3\\
\hline
136199 & Eris        & D   &         --       &         --       &         --       &         --       & 17.73 $\pm$ 0.07 & 17.49 $\pm$ 0.09\\

  \hline
\end{tabular}
\end{center}

Notes:\\
$^a$ Runs A-F as listed in table \ref{tab: obs-log}.\\
$^b$ $\Delta m_R$ is the variation in $R$-band magnitude seen for objects where (partial) light-curves were obtained. The uncertainty on each is $\sim 0.1$ mag.
\end{table*}


\begin{table*}
\caption{Average colours in $BVRIJH_S$ for all candidates (and Eris), and assessment of likely membership based on these colours.}
  \label{tab: colors}
\begin{center}
\begin{tabular}{rl cccccccc}
  \hline\hline
  \multicolumn{2}{c}{Object$^a$} & $(B-V)$ & $(V-R)$ & $(R-I)$ & $(R-J$) & $(J-H_S)$ & Vis.~slope &Ref$^b$ & Family?\\
\# & Designation & (mag.) & (mag.) & (mag.) & (mag.) & (mag.) & (\%/100nm) & & \\
  \hline
 24835 & 1995 SM 55$^\dag$  &  0.65 $\pm$ 0.01	&  0.39 $\pm$ 0.01 		&  0.36 $\pm$ 0.02 		&  0.65 $\pm$ 0.03 & -- & 2.0 $\pm$  0.8		& 1-8 & Y \\
            & 1996 RQ 20  &   0.96 $\pm$ 0.13	&  0.46 $\pm$ 0.05  		&  0.71 $\pm$ 0.12 & 	--	&                  	--		&  22.4 $\pm$  6.8 & 9,10  & N \\
 19308 & 1996 TO 66$^\dag$   & 0.68  $\pm$ 0.02 & 0.39  $\pm$ 0.01 & 0.37  $\pm$ 0.02 & 0.61 $\pm$ 0.10 & -- & 2.9 $\pm$  0.5 & 1,10-14 & Y \\
          & 1998 HL 151 &     0.67  $\pm$ 0.18 & 0.42  $\pm$ 0.16 & 0.79 $\pm$ 0.31 & --  &          --         			&  18.1 $\pm$ 16.9 & 15,16,32 & ? \\
181855 & 1998 WT 31  & 0.76 $\pm$ 0.32 & 0.51 $\pm$ 0.25 & 0.60 $\pm$ 0.28 & -- &           --       		& 16.6 $\pm$  5.2 & 32 & N \\
         & 1999 CD 158 &  0.83 $\pm$ 0.06 & 0.51 $\pm$ 0.05 & 0.54 $\pm$ 0.06 & 1.38 $\pm$ 0.09 & 		--			&  15.8 $\pm$  0.6 & 6,32 & N \\
 40314 & 1999 KR 16  &     1.07 $\pm$ 0.03 		&  0.75 $\pm$ 0.02 		&  0.74  $\pm$ 0.02 & 1.56 $\pm$ 0.08$^c$ &  0.56 $\pm$ 0.13	& 40.9 $\pm$  6.2 & 9,16-18,32 & N \\
         & 1999 OH 4   &  2.99 $\pm$ 0.48 & 0.21 $\pm$ 0.20 & 0.44 $\pm$ 0.47	& -- &		--		& 20.2 $\pm$ 35.6 & 32 & ? \\ 
 86047 & 1999 OY 3$^\dag$   &      0.75 $\pm$ 0.03 		&  0.26 $\pm$ 0.03 		& 0.33 $\pm$ 0.04 & 0.80 $\pm$ 0.12 &	-0.26 $\pm$ 0.36	& -0.5 $\pm$  5.3 & 3,32 & Y \\
86177 & 1999 RY 215 & -- & -- & -- & 0.99 $\pm$ 0.18 &	 0.52 $\pm$ 0.22 &-- & 28   & N \\
         & 2000 CG 105 & 1.11 $\pm$ 0.25   & 0.39 $\pm$ 0.13 & 0.21 $\pm$ 0.22 & -- & -- 					& 6.7 $\pm$ 17.5 & 32 & ? \\
            & 2001 FU 172 & 2.91 $\pm$ 1.53 & 0.66 $\pm$ 0.19 & 0.53 $\pm$ 0.22 & -- & --					& 39.8 $\pm$ 27.9 &32 & N \\
        & 2001 QC 298 &   0.66 $\pm$ 0.07 		&  0.37 $\pm$ 0.07 		&  0.63 $\pm$ 0.07 		&  1.06 $\pm$ 0.21 		&  0.52 $\pm$ 0.14  &  9.7 $\pm$ 10.0 &19,32 & N \\
 55565 & 2002 AW 197 &    0.93 $\pm$ 0.03 		&  0.62 $\pm$ 0.02 		&  0.55 $\pm$ 0.02 		&  1.16 $\pm$ 0.04 		&  0.39 $\pm$ 0.08 		&  22.8 $\pm$  3.5 & 20-22,32 & N \\
        & 2002 GH 32  &    0.91 $\pm$ 0.06 		&  0.66 $\pm$ 0.06 		&  0.56 $\pm$ 0.05 		&   	--	&                  	--		&  24.8 $\pm$  4.7 & 19,23 & N \\
 55636 & 2002 TX 300$^\dag$ &   0.66 $\pm$ 0.02 		&  0.36 $\pm$ 0.02 		&  0.32 $\pm$ 0.03	&   	--	& -0.47 $\pm$ 0.13 &  0.2 $\pm$  1.1 & 21,23,32 & Y \\
136108 & Haumea$^\dag$      & 0.64 $\pm$ 0.01 		&  0.33 $\pm$ 0.01 		&  0.34 $\pm$ 0.01 		& 0.88 $\pm$ 0.01 		& -0.60 $\pm$ 0.11 		& -0.6 $\pm$  0.9 & 21,24,25,32 & Y \\
         & 2003 HX 56  & 1.27 $\pm$ 1.37 & -0.26 $\pm$ 2.07 & 1.28 $\pm$ 2.10 &  -- & --	& 18.4 $\pm$ 32.7 & 32 & ? \\
120178 & 2003 OP 32$^\dag$  &    0.70 $\pm$ 0.05 &   	0.39 $\pm$ 0.06	&   0.37 $\pm$ 0.05 &   	--	& -0.51 $\pm$ 0.08 &  3.4 $\pm$  1.1 & 9,26,32 & Y \\
          & 2003 SQ 317 & 		--			&  -- & -- & 0.43 $\pm$ 0.04  & -0.45 $\pm$ 0.20 & --  & 32 & Y \\
        & 2003 TH 58  & 0.58 $\pm$ 0.12 & 0.29 $\pm$ 0.13 & 0.59 $\pm$ 0.15 & -0.13 $\pm$ 0.15 &  1.29 $\pm$ 0.20 &  3.6 $\pm$ 11.4 & 32 & N \\
        & 2003 UZ 117$^\dag$ & 		--		& -- &			--		& 	--	& -0.62 $\pm$ 0.12 & 1.1 $\pm$ 0.7$^d$ & 22,27,28,32 & Y \\  
        & 2004 PT 107 & 0.82 $\pm$ 0.21 & 0.65 $\pm$ 0.10 & 0.68 $\pm$ 0.10 & 1.15 $\pm$ 0.16 & 0.54 $\pm$ 0.22 &  27.9 $\pm$  8.4 & 32 & N \\
       & 2005 CB 79 &    0.73 $\pm$ 0.04 & 0.37 $\pm$ 0.05 & 0.36 $\pm$ 0.05 & 0.71 $\pm$ 0.08 &  -0.50 $\pm$ 0.17 &  3.1 $\pm$  2.5 & 32 & Y \\
136472 & Makemake    &   0.83 $\pm$ 0.02 		&  0.5 $\pm$ 0.1$^e$ 		&  0.3 $\pm$ 0.1$^e$ 		&   	--	&                  	--		& 7.7 $\pm$  8.8 &  21,29  & N \\
        & 2005 GE 187 & 			--		&   -- 	& --  & 	1.22 $\pm$ 0.19				& 0.65 $\pm$ 0.14 & -- & 32 & N \\
145453 & 2005 RR 43$^\dag$  & 0.77 $\pm$ 0.06 	& 0.41 $\pm$ 0.04 & 0.29 $\pm$ 0.08 & 0.48 $\pm$ 0.04 	& 	--	&  3.3 $\pm$  6.0 & 8,22,26  & Y \\
202421 & 2005 UQ 513 & --	&   	--	& --  & 1.39 $\pm$ 0.08			& 0.30 $\pm$ 0.12 & 18.1 $\pm$ 2.0$^d$ & 27,28,30,32 & N \\
\hline
136199 & Eris            &  0.78 $\pm$ 0.01 		& 0.45 $\pm$ 0.03		 & 0.33 $\pm$ 0.02 		& 0.52 $\pm$ 0.02 		& 0.25 $\pm$ 0.11 		& 5.9 $\pm$  5.6 &  21,22,31,32 & -- \\

  \hline
\end{tabular}
\end{center}

Notes:\\
$^a$ There are no published colours for candidates 1996 TR$_{66}$, 1997 RX$_{9}$, 1999 OK$_{4}$, 2000 JG$_{81}$, 2003 HA$_{57}$. 
For 2003 QX$_{91}$ and 2004 SB$_{60}$ we measured $R$-band photometry, but not colours. None of these candidates are included in the table.\\
$^b$  References:
 [1] \citet{2001-AA-378-Boehnhardt};
 [2] \citet{2001-Icarus-152-Gil-Hutton};
 [3] \citet{2002-AJ-124-Doressoundiram};
 [4] \citet{2003-Icarus-161-McBride};
 [5] \citet{2003-Icarus-161-Tegler};
 [6] \citet{2004-AA-417-Delsanti};
 [7] \citet{2007-AJ-134-Doressoundiram};
 [8] \citet{2008-AJ-136-Rabinowitz};
 [9] \citet{2001-AJ-122-Jewitt};
[10] \citet{1998-Nature-392-Tegler};
[11] \citet{1998-AJ-115-Jewitt};
[12] \citep{1999-Icarus-142-Barucci};
[13] \citep{2000-Icarus-146-Davies};
[14] \citep{2000-AA-356-Hainaut};
[15] \citep{2002-AA-389-Hainaut};
[16] \citep{2002-ApJ-566-Trujillo};
[17] \citep{2002-AJ-124-Sheppard};
[18] \citet{2006-AJ-131-Delsanti};
[19] \citep{2009-AA-494-Santos-Sanz}; 
[20] \citep{2005-PSS-53-Doressoundiram};
[21] \citep{2007-AJ-133-Rabinowitz};
[22] \citep{2009-AA-493-DeMeo};
[23] \citep{2005-Icarus-174-Doressoundiram};
[24] \citep{2006-ApJ-639-Rabinowitz};
[25] \citep{2008-AJ-135-Lacerda};
[26] \citep{2009-AA-subm-Perna};
[27] \citep{2007-AA-468-Pinilla-Alonso}
[28] \citep{2008-AA-487-Alvarez-Candal}
[29] \citep{2007-AA-468-Ortiz};
[30] \citep{2009-AA-inpress-Fornasier}
[31] \citep{2005-ApJ-635-Brown};
[32] This work.
Where colours for a given object are published by multiple authors, we quote a weighted mean.\\
$^c$ ($R-J$) calculated from near simultaneous $R$ and $J$ observations by [17] and [18] respectively. No correction is made to this (or any other colour in the table) for possible differences due to changes in rotational phase, although [17] show 1999 KR$_{16}$ to have a light-curve amplitude of $\Delta m = 0.18$ mag.\\
$^d$ Although no photometry is published, measurements of the spectral slope for these objects (derived from optical spectra) can be found in the table 2 of \citet{2009-AA-inpress-Fornasier}. We give a weighted mean for each object, and list the references to the original papers in the table.\\
$^e$ The colours for Makemake are calculated from the $BVI$ photometry from [21] along with the $R$-band photometry from [26]. 
We use a phase function of $\beta=0.05$ mag deg$^{-1}$ to correct the $R$-band photometry to zero phase angle, 
as [21] show that $\beta$ is approximately constant at this value between the $V$ and $I$ bands. The uncertainty is dominated by the
uncertainty on the $R$-band photometry.\\
$^\dag$ These objects listed as confirmed family members by \citet{2007-AJ-134-Ragozzine}.\\

\end{table*}



\begin{figure*} 
   \centering
   \includegraphics[angle=-90,width=\textwidth]{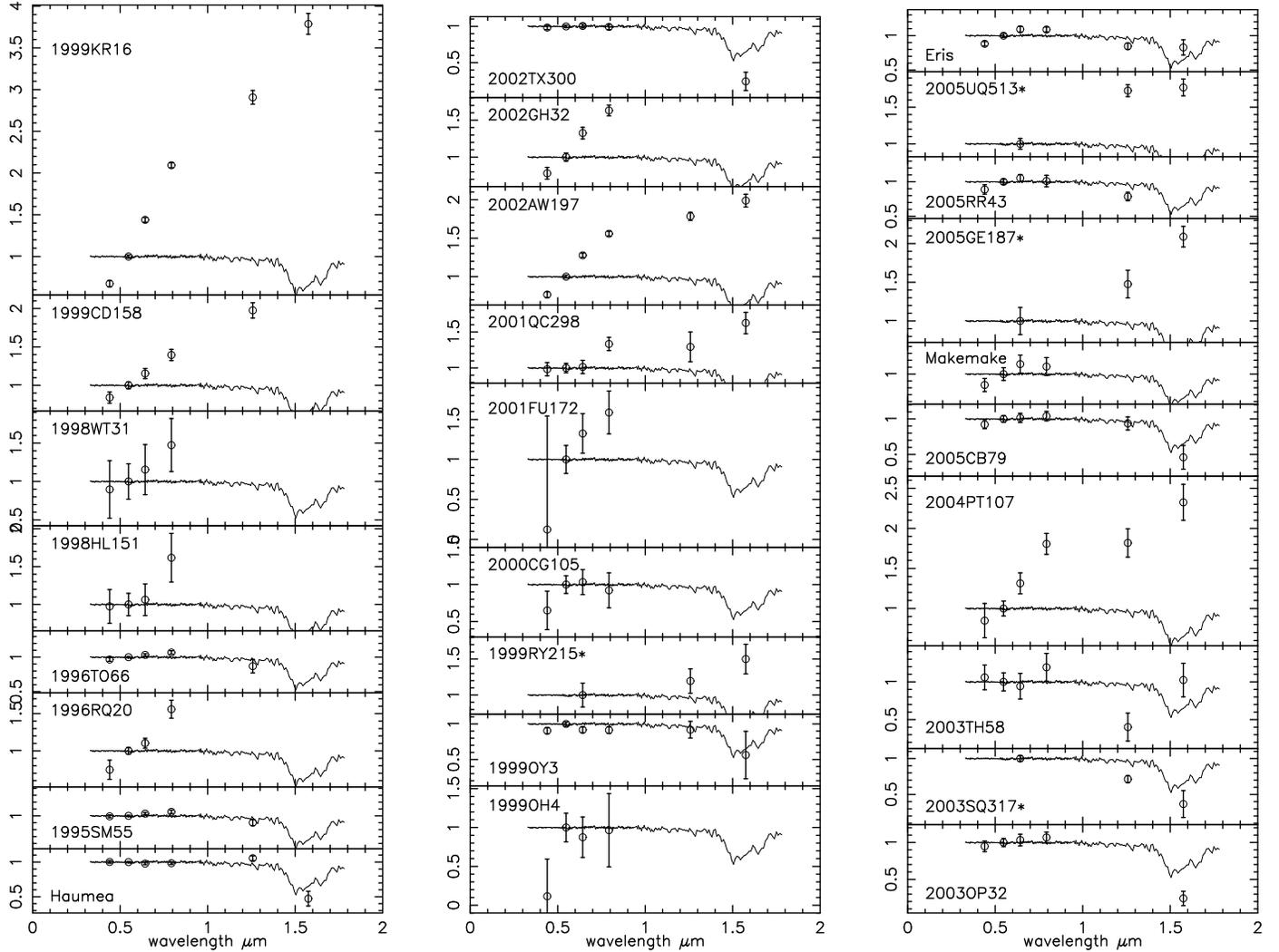} 
   \caption{Visible and NIR photometry for all candidate family members with observations in at least three bands. The data are normalized at 0.55$\mu$m ($V$ filter), except in the four cases where no $V$-band photometry exists. These data are normalised to the $R$-band, and are indicated by an asterisk next to the designation. The spectrum of Haumea is shown for comparison in each; this is taken from \citet{2009-AA-496-Pinilla-Alonso}. The photometry of the large TNO Eris is also shown for comparison; it is not associated with the family, and has a spectrum dominated by methane ice \citep{2007-AA-471-Dumas}.}
   \label{fig:SEDs}
\end{figure*}


We report the resulting photometry in table \ref{tab: phot}, where we give the mean magnitude in each band at each epoch and also an indication of the variation seen in the $R$-band where we obtained light-curves. In table \ref{tab: colors} we give the average colours of all family members that have published photometry, including our own results, taking a weighted mean where multiple measurements exist. From these average colours we calculate reflectances by comparing them to the Solar colours. To calculate the reflectance in the $H_S$ band we used a theoretical $(J - H_S)$ colour for the Sun generated by convolving the response of these filters with the Solar spectrum. We subsequently confirmed this value by observing a Solar analogue star with Hawk-I: The theoretical $(J-H_S)_\odot = 0.273$, while the value measured for the Solar twin S966 (taken from the catalogue of Solar twins in M67 by \citet{2008-AA-489-Pasquini}) is $(J-H_S)_\odot = 0.288 \pm 0.007$. These are consistent at the level of the uncertainty on our TNO colour measurements. 
We also report the visible slope for each object (\%/100 nm) in table \ref{tab: colors}, calculated from the reflectances via a linear regression over the full $BVRI$ range when it is available, or whichever measurements exist in other cases. 

The reflectance `spectra' of the TNOs from this photometry are shown in Fig.~\ref{fig:SEDs}, for all objects with photometry in at least three bands. The combined visible and NIR spectrum of Haumea from \citet{2009-AA-496-Pinilla-Alonso} is shown for comparison to the photometry. The large TNOs Eris (not a family member; observed for comparison) and Makemake (dynamically a family member candidate) are known to have methane ice surfaces from NIR spectroscopy \citep{2007-AA-471-Dumas,2007-AJ-133-Brown} and clearly differ from the Haumea spectrum. Note that those objects marked with an asterisk in the figure have their reflectance normalised to the $R$-band, as no $V$-band photometry was available. For Haumea-like neutral spectra this makes no difference, but this could give an offset in the case of red slopes; these four spectra should not be directly compared with the others in the figure, but can be compared with the Haumea spectrum.


\section{Discussion}
\label{sec: discussion}

\subsection{Family membership}

\begin{figure}
   \centering
   \includegraphics[angle=-90,width=0.45\textwidth]{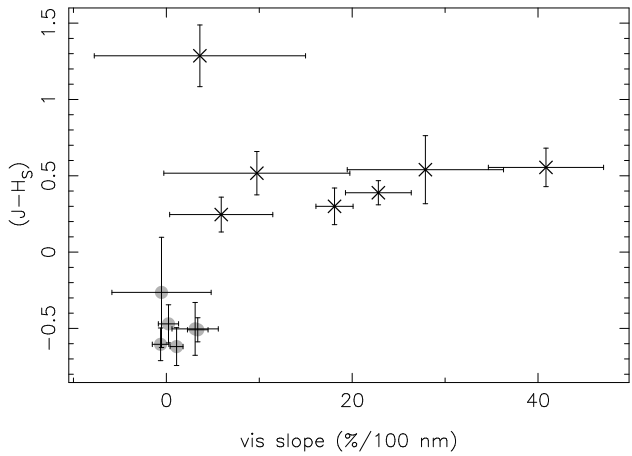} 
   \caption{$(J-H_S)$ colour against visible slope (\%/100nm) for all candidates (and Eris) and where both measurements have been made. Filled circles are confirmed family members, crosses show rejected candidates. Haumea itself is the point in the very bottom left.}
   \label{fig:JH_vis}
\end{figure}

We first wish to determine which candidates are actually family members, and which are dynamical interlopers with different surface properties. We find that the $(J-H_S)$ colour is a good diagnostic of the presence or absence of the water ice absorption feature at 1.6 $\mu$m, as expected: For Haumea we measure $(J-H_S) = -0.60 \pm 0.11$, and the colour is also significantly negative for the other known family members observed, while for the methane ice dominated comparison TNO Eris we find $(J-H_S) = 0.25 \pm 0.11$. The colours for all objects are given in table \ref{tab: colors}, along with the visible slopes, and these are also plotted in Fig.~\ref{fig:JH_vis}. In the figure there is a clear separation between the family members with negative $(J-H_S)$ at the bottom and the other objects at the top, and also a tendency for those with water ice to have blue/neutral surfaces (shallower slopes). While those without water ice have a large range of slopes from neutral to very red, there are no bodies in the lower right of the figure (water ice and red slope). We use this separation to make a rough assessment of the family membership for candidates with only optical colours; we can rule out membership for objects with very red slopes, but cannot use a blue slope to confirm membership.

We confirm two more family members in addition to those listed by \citet{2007-AJ-134-Ragozzine}; 2003 SQ$_{317}$ and 2005 CB$_{79}$. These have $(J-H_S) = -0.45 \pm 0.20$ and $-0.50 \pm 0.17$ respectively. 2005 CB$_{79}$ has since been confirmed as a family member by NIR spectroscopy \citep{2008-ApJ-684-Schaller}. For 2003 SQ$_{317}$ the lack of optical colours as supporting evidence and the relatively large uncertainty on $(J-H_S)$ makes the water ice detection preliminary, and spectroscopy or further photometry would be worthwhile, but the evidence is certainly as strong as for some previous spectroscopic water ice `detections' so we choose to regard this as a confirmed family member for the purposes of this paper. This brings the total number of confirmed family members to 10, of the 35 candidate objects. We are far more efficient at rejecting candidates though; 8 objects have $(J-H_S)$ colours inconsistent with water ice, and cannot be true family members. These are 1999 KR$_{16}$, 1999 RY$_{215}$, 2001 QC$_{298}$, 2002 AW$_{197}$, 2003 TH$_{58}$, 2004 PT$_{107}$, 2005 GE$_{187}$ and 2005 UQ$_{513}$. This is in agreement with \citet{2008-AA-489-Pinilla-Alonso}, who rejected 2005 UQ$_{513}$ on the basis of a very red slope in an optical spectrum. We also find that 1998 WT$_{31}$ and 2001 FU$_{172}$ have strongly red visible slopes, and can probably be rejected as family members without Hawk-I data. Including also Makemake and 2004 SB$_{60}$, which have been shown to lack water ice on their surfaces by NIR spectroscopy \citep{2007-AJ-133-Brown,2008-ApJ-684-Schaller} and the others listed in the introduction which have previously been found to have very red optical colours, this gives a total of 15 of the 35 candidates that are shown not to belong to the family.  Finally, we also observed 1998 HL$_{151}$, 1999 OH$_{4}$, 2000 CG$_{105}$ and 2003 HX$_{56}$ in the optical, but all of these were too faint to put meaningful constraints on their family membership. We summarise which objects we believe to be family members, which we can rule out, and which we do not yet have enough information on in the last column of table \ref{tab: colors}.

Given the high rate of rejection of candidates, we consider the likelihood that this is a true family from a statistical point of view, or whether the $\sim 30\%$ of water ice bodies within the candidates could just reflect the proportion within the TN region in general. Based on the TNO taxonomy proposed by \citet{2008ssbn.book..181F}, the confirmed family members all belong to the BB class, while the rejected candidates come from all 4 of the groups (the majority of the newly rejected ones are from the red classes RR and IR, since they were mostly rejected due to their red slopes). The BB class makes up only 20\% of the whole TNO population; the proportion of BB within the candidates ($\equiv$ confirmed family members / candidates $ \approx 30\%$) is high but not so unusual given the small numbers of objects involved. If instead of taxonomic classes we consider the proportion of TNOs with water ice detections (from IR spectroscopy), then in the case of the general population we find $\sim 50\%$ (from table 1 of \citet{2008ssbn.book..143B}), making the proportion of bodies with water ice in the candidates lower than the general population, although this number contains significant biases as the spectroscopy only covers the brightest bodies. We can conclude that we do not see a significantly larger number of water ice bodies in the candidate list than in the general population, but this ignores grouping in orbital element space.

\begin{figure}
   \centering
   \includegraphics[width=0.45\textwidth]{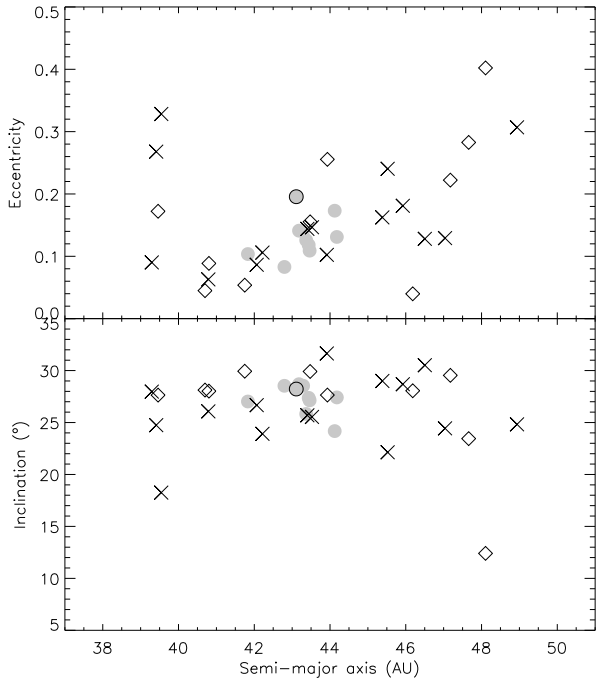} 
   \caption{Confirmed family members (grey filled circles), rejected candidates (crosses) and those with unknown surface properties (open diamonds) plotted in terms of the orbital parameters semi-major axis, inclination and eccentricity. Haumea itself is shown as a grey circle with a black outline.}
   \label{fig:dynamics}
\end{figure}

In Fig.~\ref{fig:dynamics} we show the candidates in terms of their orbital parameters semi-major axis, inclination and eccentricity. The confirmed family members cluster tightly around the centre of the distribution in both plots, where the original orbit of the pre-collision Haumea was (Haumea itself now has a higher eccentricity than the centre of the family due to interaction with Neptune through orbital resonance \citep[see ][]{2007-AJ-134-Ragozzine}). This suggests that the family hypothesis is a valid one, but that the spread in orbital elements since the collision is less than the range investigated by \citet{2007-AJ-134-Ragozzine}. Taking the required collision velocities from that paper ($\delta v_{\rm min}$; the minimum ejection velocity required including the effects of eccentricity and inclination diffusion in mean-motion resonances) we find that the largest velocity required by any confirmed family member is 123.3 m/s (for 1995 SM$_{55}$), while candidates are listed with $\delta v_{\rm min}$ up to 250 m/s. If we restrict the candidate list to those with $\delta v_{\rm min} \le 150$ m/s, we find that the proportion of confirmed water ice detections rises to 53\%, and goes up to 64\% if we look only at those with lower $\delta v_{\rm min}$ than 1995 SM$_{55}$, so the grouping is statistically significant compared with the general population of bodies with water ice surfaces within the TN region. It should be noted though that there are rejected candidates spread evenly across the phase space, including 2005 UQ$_{513}$ and 1999 CD$_{158}$ near to the centre of the family distribution, which demonstrates the importance of physical studies of the candidates to confirm membership. The remaining unknown objects near to the centre of the distribution are 1999 OK$_{4}$ and 2003 QX$_{91}$ (although the latter has high eccentricity and a high $\delta v_{\rm min}$ of 222 m/s) which should be high priority targets for further study to measure candidate family member surface properties, along with 1997 RX$_{9}$ which has a low $\delta v_{\rm min}$ of 86.8 m/s.

It is noticeable that the confirmed members remain the larger bodies, even though this photometric method is sensitive to water ice absorption on bodies too small for NIR spectroscopy. We tested the idea that retention of a water ice surface could be a property of only the larger TNOs by looking for a correlation between absolute magnitude and the $(J-H_S)$ index, but found that no such correlation exists. It is likely that there are smaller water ice covered family members, however they have yet to be discovered or confirmed. We also tested for any correlation of the colour with orbital elements and found none; we are dealing with a family clustered in dynamical element space, not a consequence of any correlation of, for example, the presence of water ice with semi-major axis.

\subsection{Light-curves}

This work aimed to test both the membership of the candidate family members and also the hypothesis that the family members apart from Haumea itself are composed of almost pure water ice, being made of the reassembled fragments of the outer layers of the differentiated proto-Haumea \citep{2007-Nature-446-Brown}. This can be probed by testing the density of the family members; Haumea is known to have a rock-like density of 2.5-3.3 g cm$^{-3}$ (the value found from combining the size from \citet{2008ssbn.book..161S} and mass from \citet{2009-AJ-137-Ragozzine} agrees with the value from the light-curve model of \citet{2006-ApJ-639-Rabinowitz}) but the other family members should have densities at or below the density of water ice, $\sim 1$ g cm$^{-3}$. To test this we sought to apply the technique of measuring rotation rate and elongations using light-curves, which then constrain the density of a strengthless body to be
\begin{equation}
\rho \ge \frac{10.9}{P_{\rm rot}^2}\frac{a}{b} \hspace{5pt} \textrm{g cm}^{-3},
\end{equation}
where $a/b$ is the axial ratio for an ellipsoid and the rotation period $P_{\rm rot}$ is in hours \citep{2000-Icarus-148-Pravec}. We can reasonably expect the recombined fragments from a collision to be a loosely packed `rubble pile' and therefore strengthless. This method only gives a lower limit to the density as the object does not need to be spinning at its break up velocity, but must be below it, and also that the light-curve amplitude $\Delta m$ only gives a lower limit on the elongation, $a/b \ge 10^{0.4\Delta m}$. Despite this, when studying a population a cut off in minimum densities becomes apparent, which can be used as a reasonable measurement of the bulk density of the bodies in the population. This has been clearly demonstrated for asteroids \citep{2002-AsteroidsIII-2.2-Pravec}, where there are many ($N > 1000$) light-curves available, and also used to derive a low bulk density for cometary nuclei \citep{2006-MNRAS-373-Snodgrass}, in agreement with the results found by the \textit{Deep Impact} mission \citep{2007-Icarus-190-Richardson}, despite the relatively low number of light-curves available for nuclei.

We measured partial light-curves for 13 of the candidates using EFOSC2, however poor weather during these runs prevented us from building up the number of light-curves required to study the density of these bodies by this statistical technique. This was further hampered by the large proportion of the candidates which were eventually rejected as non-family members. For 1998 WT$_{31}$ and 1999 OY$_{3}$ we have less than 10 points spread over three and two nights respectively and there is no significant variation. 1999 RY$_{215}$ and 2003 TH$_{58}$ also show no significant variation despite larger data sets. For 1999 CD$_{158}$, 2000 CG$_{105}$, 2002 GH$_{32}$ and 2005 UQ$_{513}$ the light-curves show significant variation, with ranges of $\Delta m = 0.6$, 0.45, 0.75 and 0.3 mag., but no period could be determined. For 2001 QC$_{298}$ there is possibly a maximum each night in the data, with $\Delta m = 0.4$, but there can be other periods beyond the $\sim 12$ hours best fit. 
2003 SQ$_{317}$ gives a good fit with single peak light-curve of 3.7 hours, while a double peaked light-curve at 7.5 hours also looks reasonable. 2004 PT$_{107}$ shows a possible slight variation (0.05 mag), but not a very convincing one, with a suggested long period ($\sim 20$ hours). We obtained data on 2004 SB$_{60}$ on two nights which show a variation of $\Delta m=0.2$, but no clear periodicity. There is a possible solution at around 17.5 hours, but it is not convincing.
2005 GE$_{187}$ has a reasonably convincing single peak light-curve with a 6.1 hour period and $\Delta m = 0.5$. 

\begin{figure}
   \centering
   \includegraphics[angle=-90,width=0.45\textwidth]{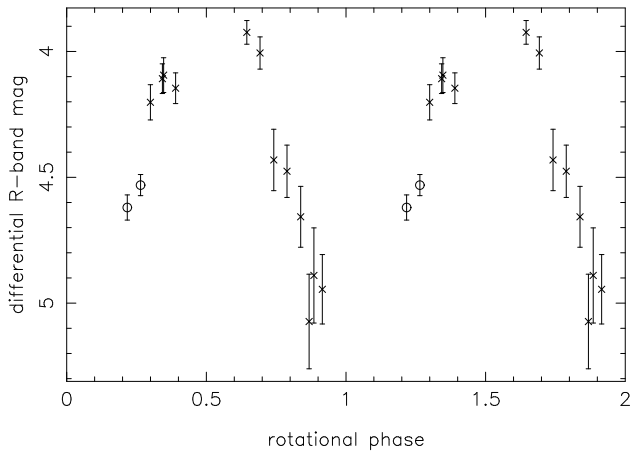} 
   \caption{Light-curve for 2003 SQ$_{317}$, with data taken on two nights (circles from the night of August 30th 2008 and crosses from August 31st) folded onto a 3.74 hour period.}
   \label{fig:2003SQ317-lc}
\end{figure}

The only light-curve in this set of relevance to the density of the family members is that of 2003 SQ$_{317}$, shown in Fig.~\ref{fig:2003SQ317-lc}. The period of 3.7 hours and the range of $\Delta m=1.0$ mag implies a high density, $\rho \ge 2.0$ g cm$^{-3}$, however this is for a single peaked light-curve as would be caused by albedo variations and not shape. The light-curves of Solar System minor bodies are more likely to be caused by shape than albedo patterns \citep{2008-SBPS-Jewitt,2008ssbn.book..129S}. Assuming that this single peak light-curve shows half of the period of the true shape controlled light-curve, the implied density is $\rho \ge 0.5$ g cm$^{-3}$, which is a weak constraint. Given the sparse light-curve coverage there are also other possible periods. We cannot rule out a low density and therefore an entirely ice composition for this body. 

Of the other family members 5 of the large bodies with confirmed water ice surfaces also have light-curves (not including Haumea itself). 1995 SM$_{55}$ has a rotation period of 8.08 hours and $\Delta m=0.19$ \citep{2003-EMP-92-Sheppard} (implying $\rho \ge 0.20$ g cm$^{-3}$). 1996 TO$_{66}$ has $P_{\rm rot}=7.9$ hours, $\Delta m = 0.26$ \citep{2003-EMP-92-Sheppard} ($\rho \ge 0.22$), but both the period and amplitude of the light-curve are seen to change \citep{2000-AA-356-Hainaut}. 2002 TX$_{300}$ has a period between 8 and 12 hours and a low amplitude of $\Delta m =0.08$ \citep{2003-EMP-92-Sheppard} ($\rho \ge 0.18$). \citet{2009-AA-submitted-Perna} find that 2005 RR$_{43}$ has $P_{\rm rot}=5.08$ hours, $\Delta m = 0.12$  ($\rho \ge 0.47$). Observations from the same group find no obvious periodicity for 2003 UZ$_{117}$. None of these light-curves require high densities, although for these very large objects it is also likely that the rubble pile assumption will be invalidated due to compaction by self gravity, in which case finding the density from the light-curve involves assuming fluid like behaviour \citep[see][]{2007-AJ-133-Lacerda}.


\section{Summary}

We have presented optical and/or near infrared colours for 22 of the 35 candidate members of Haumea's collisional family that were listed by \citet{2007-AJ-134-Ragozzine}. We make use of a unique capability of the new Hawk-I instrument at the VLT to evaluate the depth of the 1.6 $\mu$m water ice absorption band using NIR photometry on objects too faint for spectroscopy. We find:

\begin{enumerate}
\item Of the 15 candidates observed with Hawk-I, 7 were found to be family members. Most (6) of these were already known family members, including Haumea itself, whose confirmation proves the validity of the photometric technique used. In addition to the confirmed family members listed by \citet{2007-AJ-134-Ragozzine} we confirm the identification by \citet{2008-ApJ-684-Schaller} of water ice on 2005 CB$_{79}$, and identify 2003 SQ$_{317}$ and as a probable new family member.

\item We reject the other 8 candidates observed with Hawk-I as interlopers which lack water ice absorption. In general the rejected bodies are relatively far from the centre of the family in orbital parameter space.

\item We present optical colours for 10 candidates and also collect all available colour information from the literature for the full set. Of the 20 candidates not yet observed with Hawk-I there are optical colours for 13. We find that all objects where the NIR colour indicates water ice have neutral or blue slopes, and consequently we can reject the possibility of water ice on the surface of the very red objects in this sample with a reasonable degree of confidence. In this way we rule out family membership for a further 5 of the candidates, in addition to 2 candidates which are already known to have no water ice on their surface from NIR spectroscopy.

\item Of the 35 family member candidates this gives totals of 10 confirmed members (29\%), 15 non-members (43\%) and 10 that still have to have their surfaces characterised. It appears that the family members all fall within the centre of the dynamical region searched by \citet{2007-AJ-134-Ragozzine}, so we expect that most of the remaining bodies will also be rejected.

\item We obtained partial $R$-band light-curves for 13 of the candidates, only two of which were subsequently confirmed as a family members. Of these 1999 OY$_{3}$ showed no significant variation in the short sequence we were able to obtain on it, while 2003 SQ$_{317}$ shows variations consistent with a 3.74 hour single peak light-curve, but other periods are possible in the sparse data. Neither this nor the existing light-curves in the literature for other family members provide strong constraints on the density of these bodies, so we cannot yet determine whether or not they are `pure' water ice bodies formed from the outer layers of the pre-collision Haumea.

\end{enumerate}


\begin{acknowledgements}
We thank the dedicated staff of ESO's La Silla and Paranal observatories for their assistance in obtaining this data, and in particular Giovanni Carraro for providing us with the Hawk-I observations of the Solar analogue. We are grateful to Noemi Pinilla-Alonso for providing the spectrum of Haumea and Davide Perna for providing us with his results in advance of their publication. We thank Pedro Lacerda and Franck Marchis for helpful suggestions. We also thank the referee, David Rabinowitz, for constructive comments that improved the paper.
\end{acknowledgements}

\bibliographystyle{aa}      
\bibliography{biblio}

\end{document}